# Effect of Covalent Functionalisation on Thermal Transport Across Graphene-Polymer Interfaces

Y. Wang[1], H. F. Zhan[2], Y. Xiang[1], C. Yang[1], C. M. Wang[3], Y. Y. Zhang[1*]

[1]School of Computing, Engineering and Mathematics, University of Western Sydney, Locked Bag 1797, Sydney NSW 2751, Australia
[2]School of Chemistry, Physics and Mechanical Engineering, Queensland University of Technology, 2 George St, Brisbane QLD 4001, Australia
[3]Engineering Science Programme, Faculty of Engineering, National University of Singapore, Kent Ridge, Singapore 119260

*Corresponding Author: Tel.:+61 2 47360606. Fax +61 2 47360833. Email: yingyan.zhang@uws.edu.au

**ABSTRACT:** This paper is concerned with the interfacial thermal resistance for polymer composites reinforced by various covalently functionalised graphene. By using molecular dynamics simulations, the obtained results show that the covalent functionalisation in graphene plays a significant role in reducing the graphene-paraffin interfacial thermal resistance. This reduction is dependent on the coverage and type of functional groups. Among the various functional groups, butyl is found to be the most effective in reducing the interfacial thermal resistance, followed by methyl, phenyl and formyl. The other functional groups under consideration such as carboxyl, hydroxyl and amines are found to produce negligible reduction in the interfacial thermal resistance. For multilayer graphene with a layer number up to four, the interfacial thermal resistance is insensitive to the layer number. The effects of the different functional groups and the layer number on the interfacial thermal resistance are also elaborated using the vibrational density of states of the graphene and the paraffin matrix. The present findings provide useful guidelines in the application of functionalised graphene for practical thermal management.

# 1. INTRODUCTION

As the modern electronic industry moves forward with high power consumption, integrated functions and miniaturisation, the power densities of modern electronic devices have been increased dramatically. Thus, the dissipation of large amounts of heat has become a critical issue in the electronics industry and this problem has generated great research interests.

An enormous research effort has been carried out worldwide to improve the thermal management efficiency of electronic packages. In a modern high power integrated circuit (IC) package, thermal interface material (TIM) is the material sandwiched between two contacting solid surfaces, i.e. between the IC chip and the heat sink. The TIMs are used to eliminate the interstitial air gaps between the contacting solid surfaces, and thereby enhancing the heat transfer from the IC chip to the heat sink. TIMs are composites comprising a polymer-based material matrix and thermally conductive fillers.[1] Silver, copper, alumina and aluminium nitride are the most commonly used fillers in the current TIMs.[2,3] The thermal conductivity of current TIMs varies from 1 to 10 $Wm^{-1}K^{-1}$ with a filler volume fraction of up to 70%.[3] It is widely recognised that the thermal conductivity of TIMs is one of the main bottlenecks for the thermal management efficiency of electronic packages. To tackle this challenge, the most straightforward strategy is to use highly conductive fillers in TIMs, such as carbon nanotubes and graphene.[1-3]

It was experimentally evidenced that graphene possesses an extremely high thermal conductivity of up to 5000 $Wm^{-1}K^{-1}$ at room temperature.[4,5] This thermal conductivity is more than an order of magnitude higher than that of the traditional fillers used in present TIMs. Extensive research studies have therefore been conducted using graphene as fillers to make graphene-based TIMs. However, the reported thermal conductivity of graphene-polymer composite TIMs spreads from 1 to 7 Wm$^{-1}$K$^{-1}$, which is as low as those of the traditional TIMs.[6-18] It has been found that one of the main reasons for the unsatisfactory results is the high interfacial thermal resistance (or Kapitza resistance) between the graphene fillers and polymer matrix, which degrades the overall thermal conductivity of TIMs.[9-18] Various techniques have been proposed to reduce the graphene-polymer interfacial thermal resistance.[19-22] For example, Konatham et al.[19] investigated the interfacial thermal resistance at the graphene-octane interfaces using molecular dynamics (MD) simulations. By functionalizing the graphene with hydrocarbon chains, they found that the graphene-octane interfacial thermal resistance can be reduced by up to 50%. Wang et al.[20] conducted MD simulations to examine the thermal transport across the interfaces between graphene and polyethylene. By using $C_{15}H_{31}$ hydrocarbon chains as the functional group for the graphene, they found that the interfacial thermal conductance can be enhanced by more than two times when the coverage is increased to 0.0144 Å$^{-2}$. Ganguli et al.[21] conducted experiments to prepare composite TIMs filled with the silane functionalised graphene and pristine graphene, respectively. Their experimental measurements indicated that the thermal conductivity of TIMs filled with functionalised graphene may be improved by more than 50%, compared to that of the TIMs filled with pristine graphene. So far, the research findings have clearly proven that covalent functionalisation is indeed very efficient in reducing the graphene-polymer interfacial thermal resistance, thereby furnishing TIMs with a higher thermal conductivity. However, it is presently not clear which functional group is the most efficient one since previous research studies have only focused on graphene functionalised by one single type of functional group. In addition, the widely used TIM polymer matrix - paraffin was rarely studied.

In this study, extensive MD simulations will be carried out with the view to investigate the thermal transport across graphene-paraffin interfaces. The effects of different covalent functional groups, coverage of functionalisation, number of graphene layers in enhancing thermal transport across graphene-paraffin interfaces will be examined.

## 2. COMPUTATIONAL METHODS

The thermal transport properties of the graphene-paraffin composites were explored using the reverse non-equilibrium molecular dynamics (RNEMD) simulation method, which is based on Muller-Plathe's approach.[23] All the simulations were performed using large-scale atomic/molecular massively parallel simulator - LAMMPS.[24] The interaction between carbon atoms in graphene is described by the bond order AIREBO potential[25] since it has been widely used for simulations of graphene and producing reliable results.[26,27] The polymer consistent force field (PCFF)[28,29] is used to model the polymer molecules and various functional groups. The PCFF has been employed to investigate the thermal transport properties in polymeric materials and the obtained results have been found to be in excellent agreement with experimental results.[30-34] The interactions between graphene and polymer, or those between functional group and polymer are van der Waals (vdW) interactions, which are described by the Lenard-Jones (LJ) potential. In this study, the LJ potential parameters of the carbon atoms in graphene are adopted from previous research on the graphene-polymer interfacial thermal transport.[35] The LJ potential parameters of different atoms in polymer and functional groups are taken from Material Studio based on the PCFF. The detailed LJ potential parameters used in this study are given in Table 1. The LJ potential parameters across different species of atoms are obtained by using the Lorentz-Berthelot mixing rules ($\varepsilon_{ij} = sqrt(\varepsilon_i \varepsilon_j)$; $\sigma_{ij} = (\sigma_i + \sigma_j)/2$, where $\varepsilon$ and $\sigma$ denote the energy and distance constants respectively, subscripts $i$ and $j$ refer to different atom species).

**Table 1. LJ potential parameters of different atoms species**

| atom species | energy constant ε (eV) | distance constant σ (Å) |
|---|---|---|
| carbon (in graphene) | 0.002390 | 3.412 |
| carbon (in polymer, methyl or butyl) | 0.002342 | 4.010 |
| carbon (in phenyl) | 0.002775 | 4.010 |
| carbon (in formyl or carboxyl) | 0.005204 | 3.308 |
| oxygen (double bonded to carbon) | 0.011578 | 3.300 |
| oxygen (in hydroxyl group) | 0.010407 | 3.535 |
| Nitrogen | 0.002819 | 4.070 |
| hydrogen (bonded to carbon) | 0.000867 | 2.995 |
| hydrogen (bonded to oxygen or amine) | 0.000564 | 1.098 |

The polymer used in the present work is paraffin ($C_{30}H_{62}$), which has been widely used as the matrix of TIMs. The paraffin blocks were firstly constructed in Material Studio, and then relaxed in a canonical (NVT) ensemble (i.e. constant number of atoms, volume and temperature) in order to release residual stresses. Thereafter, a graphene-paraffin composite model was constructed by sandwiching graphene sheets between the relaxed paraffin blocks (see Figure 1(a)). All the models were designed to have a block size of 29 Å × 29 Å × 160 Å. In a previous MD simulation research, Luo and Lloyd[35] have studied the effect of model size on the interfacial thermal conductance across graphene-paraffin interfaces. It was found that there is no clear size dependence on the interfacial thermal conductance when the paraffin block length varied from 35 Å to 82 Å. It is explained that the heat is conducted in the paraffin via diffusive vibration modes, which have very short propagation lengths (of the order of a few bond lengths). Thus, a block of length 35 Å is large enough to include all the important thermal transport modes in amorphous paraffin. For the cross sectional size, Luo and Lloyd[35] have found that the graphene-paraffin interfacial thermal conductance can be enhanced by merely 15% when the cross sectional width was increased from 19.68 Å to 78.72 Å. Above 78.72 Å, such enhancement is saturated, which indicates that all the important modes are excited. In the present work, we limit our attention on the effect of different covalent functional groups. Based on the findings of Luo and Lloyd,[35] the models were designed to have a size of 29 Å × 29 Å × 160 Å. Periodic boundary conditions were applied in all three directions. A small time step of 0.25 fs was chosen for all simulations, due to the presence of light hydrogen atoms. Prior to the simulations, an energy minimisation was performed using the conjugate gradient algorithm. Thereafter, the composite model was annealed in a canonical ensemble from 300 K to 1000 K, relaxed at 1000 K for 100 ps and then cooled down to 300 K. Finally, the whole system was relaxed in an isothermal-isobaric (NPT) ensemble (i.e. constant number of atoms, pressure and temperature) at 300K and 1 atm for 500 ps.

The core idea of RNEMD simulation is to apply a heat flux to a system and determine the resultant temperature gradient. As shown in Figure 1, heat source and sink regions are defined in the middle and at the two ends of the composite model, respectively. A heat flux is injected between the heat source and sink regions by exchanging the kinetic energies between the hottest atom in the heat sink region and the coldest atom in the heat source region in a microcanonical (NVE) ensemble (i.e. constant number of atoms, volume and energy). The exchange process was performed every 1000 time steps. The heat flux $J$ due to the exchange of kinetic energies is determined by

$$J = \frac{\Sigma_{N_{transfer}} \frac{1}{2}(m_h v_h^2 - m_c v_c^2)}{2At_{transfer}} \tag{1}$$

where $t_{transfer}$ is the summation time, $N_{transfer}$ is the number of exchange, $A$ is the cross-sectional area that the heat energy passes through, subscripts $h$ and $c$ refer to the hottest and coldest atoms of which the kinetic energies are interchanged. Once a steady state is reached, a stable temperature gradient between the heat source and sink regions can be established as shown in Figure 1(b). Owing to the presence of the polymer-graphene interface, there exists a temperature jump, $\Delta T$ at the interface. The graphene-polymer interfacial thermal resistance $R_K$ is then calculated by

$$R_K = \Delta T/J \qquad (2)$$

The final interfacial thermal resistance value was calculated by averaging the data collected over 500 ps period in the steady state. The error bars were then obtained using the block averaging approach.[36]

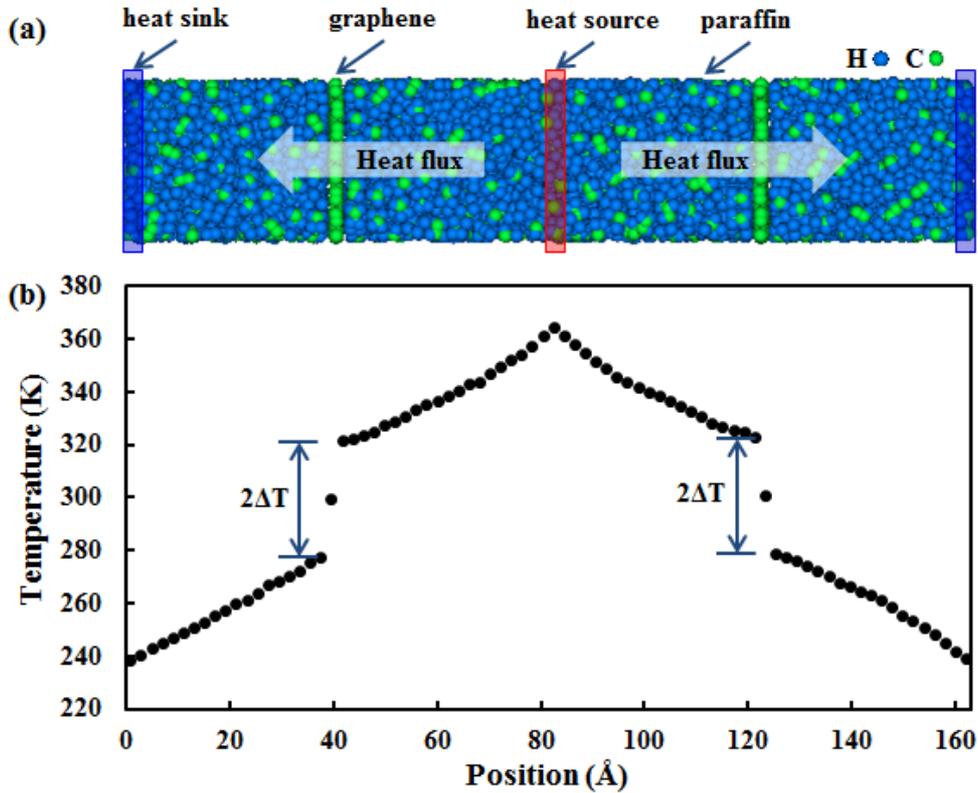

**Figure 1.** (a) Pristine graphene and paraffin composites in RNEMD simulation; and (b) the resultant temperature gradient.

## 3. RESULTS AND DISCUSSION

In order to validate the models used herein, simulations were first conducted on the pure paraffin with approximate dimensions of 29 Å × 29 Å × 74 Å. By noting that $\lambda = J/(2A\, \partial T/\partial x)$, where $\partial T/\partial x$ is the temperature gradient, the thermal conductivity $\lambda$ of pure paraffin with a density of 0.887 g cc$^{-1}$ is calculated to be 0.309 ± 0.014 Wm$^{-1}$K$^{-1}$ at 300 K. This value is in good agreement with those of pure paraffin obtained from simulations and experimental measurements in literature, i.e. 0.220-0.345 Wm$^{-1}$K$^{-1}$.[35,37,38]

### 3.1. Effect of Covalent Functionalisation

First, the thermal transport across the interface between pristine graphene and paraffin was investigated. Figure 2(a) shows the pristine graphene model of size 29 Å × 29 Å. By placing it into the paraffin as shown in Figure 1(a), a composite system was constructed with approximate dimensions of 29 Å × 29 Å × 160 Å. After performing RNEMD simulations, the interfacial thermal resistance between pristine graphene and paraffin was found to be 0.669 ± 0.043 ×10$^{-8}$ m$^2$KW$^{-1}$. These results are in good agreement with those obtained by other researchers (0.666 to 0.909 ×10$^{-8}$ m$^2$KW$^{-1}$).[30,35,39] Using the interfacial thermal resistance between pristine graphene and paraffin as a reference, the effect of functionalisation on the graphene-paraffin interfacial thermal resistance can be established. The functionalisation effect was examined through simulating the graphene with different covalent functional groups, including butyl (-C$_4$H$_9$), methyl (-CH$_3$), phenyl (-C$_6$H$_5$), formyl (-COH), carboxyl (-COOH), amines (-NH$_2$) and hydroxyl (-OH). The chosen functional groups are commonly used ones and they have been applied for covalently functionalising the graphene in either experimental or theoretical studies in the literature.[40-44] Some of the functional groups, i.e. hydroxyl and carboxyl, exist in graphene oxide. In all the cases, for generality, the functional groups were randomly distributed onto both sides of the graphene and the coverage of functionalisation ranges from 0.60% to 5.36%. Figures 2(b)-(h) show graphene with different functional groups at a coverage of 1.79%. The functional groups are bonded to the graphene as shown in Figures 2(b)-(h). In MD simulations, the interactions between the functional groups and the graphene are described by a combination of the PCFF and LJ potential. Specifically, a functional group interacts with its bonded carbon atom in graphene based on the PCFF whereas it interacts with other carbon atoms in graphene based on the LJ potential. Figure 3 displays the paraffin composite with graphene functionalised with 5.36% butyl, as well as the resultant temperature gradients. It is clearly shown in Figure 3(b) that the temperature changes linearly along the heat flux direction with a sudden temperature jump at the interface due to the different thermal properties of the graphene and the paraffin. In a previous research, Hu and Poulikakos[45] found that for some materials, i.e. GaN, that adhered to a monolayer graphene, a significant temperature jump may exist at the near-interface region due to the effect of graphene on the interface atoms. This effect was not observed in the present work. It may indicate that paraffin cannot be significantly affected by the neighboring monolayer graphene.

Based on Eq. (2), the interfacial thermal resistance $R_K$ can be obtained by MD simulations for different functionalised graphene-paraffin composites. In order to have a better insight into the effect of different functional groups, relative interfacial thermal resistance $R_K/R_{K0}$ (where $R_{K0}$ is the interfacial thermal resistance between pristine graphene and paraffin) with respect to the coverage is presented in Figure 4. It is clearly seen that different covalent functional groups affect the interfacial thermal resistance significantly. In general, the presence of functionalisation leads to a reduction of graphene-paraffin interfacial thermal resistance and this reduction is dependent on the coverage. Among the various functional groups, butyl yields the most prominent reduction. With a 5.36% coverage of butyl, the interfacial thermal resistance between graphene and paraffin is reduced by 56.5%. Besides butyl, it is found that the functionalisation using methyl, phenyl and formyl groups is also effective in enhancing the interfacial thermal transport. For instance, with a coverage of 5.36% methyl, phenyl and formyl groups, one obtains an interfacial thermal resistance reduction of 40.1%, 42.5% and 35.5%, respectively. Other functional groups, including carboxyl, hydroxyl and amines are found to be ineffective in enhancing the thermal transport across the graphene-paraffin interface. The reduction of interfacial thermal resistance induced by them is no more than 15% with a coverage of up to 5.36%.

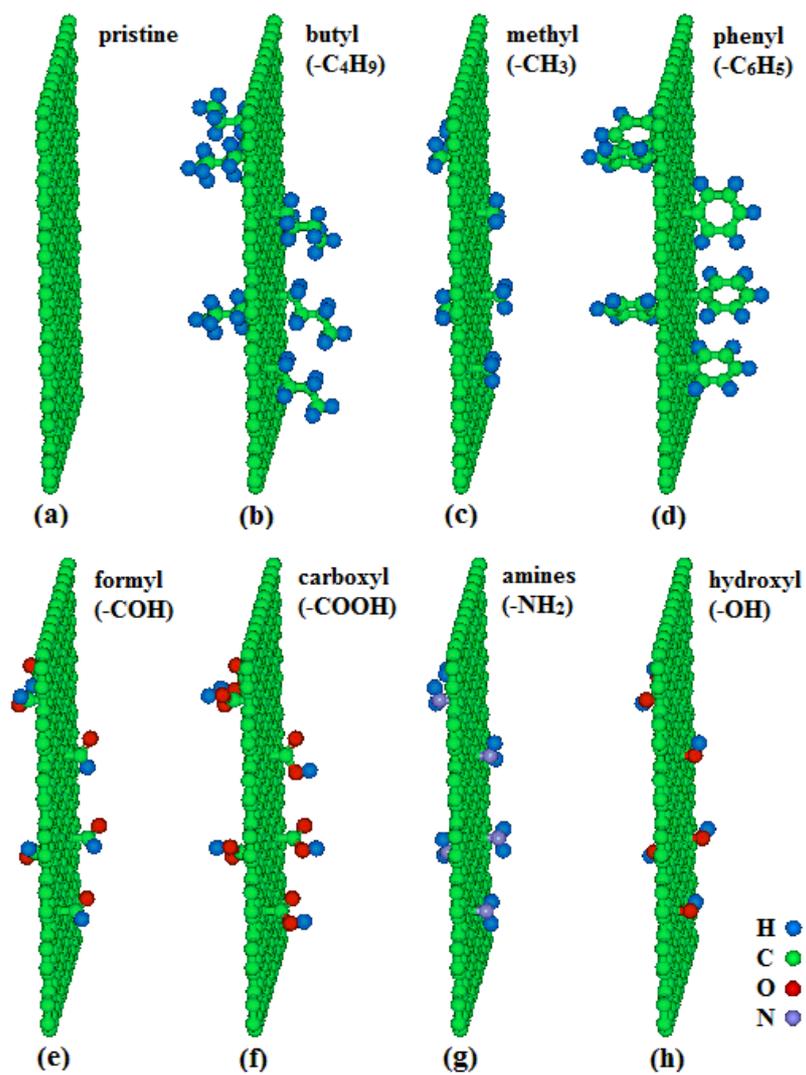

**Figure 2.** Simulation models of (a) pristine graphene and graphene functionalised with (b) butyl, (c) methyl, (d) phenyl, (e) formyl, (f) carboxyl, (g) amines and (h) hydroxyl.

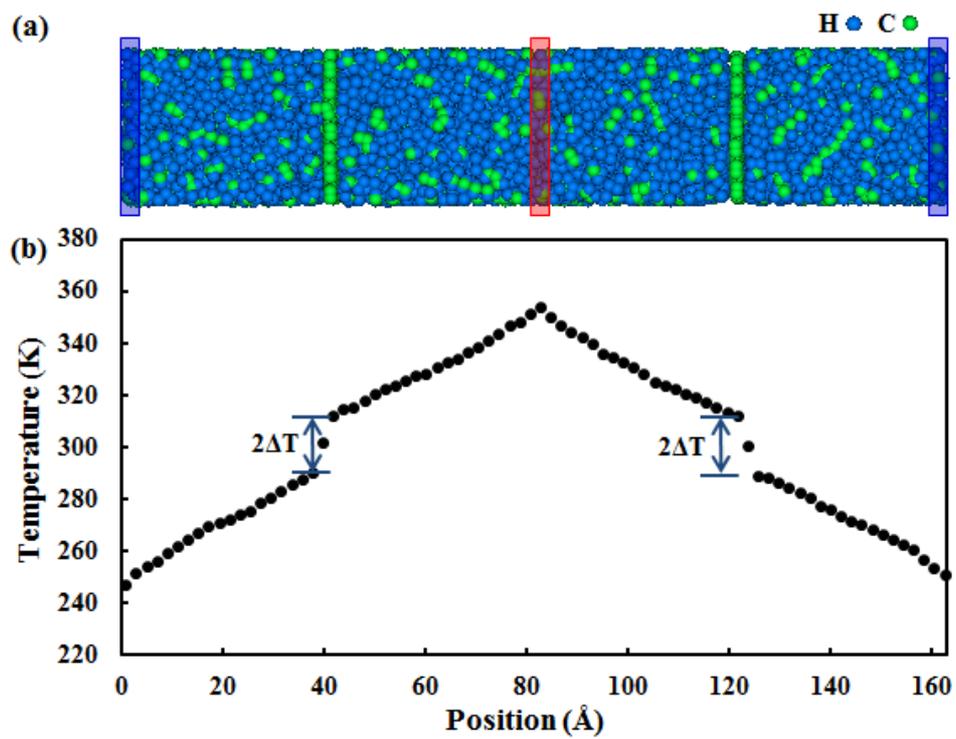

**Figure 3.** (a) 5.36% butyl (-C$_4$H$_9$) functionalised graphene and paraffin; and (b) the resultant temperature gradient.

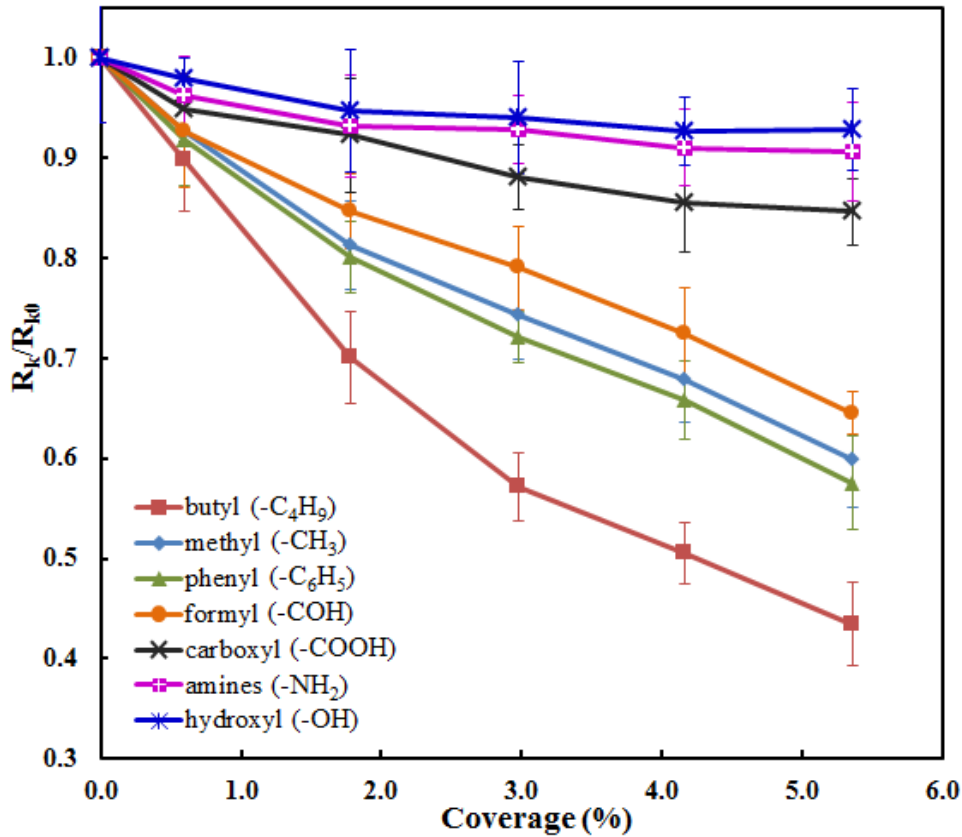

**Figure 4.** Variation of the relative interfacial thermal resistance with respect to the coverage of various types of functionalisation.

To gain a better understanding on the interfacial thermal resistance, the two major theoretical models, i.e. acoustic mismatch model (AMM) and diffuse mismatch model (DMM) were considered.[46] Both models establish that the overlap of vibrational density of states (VDOS) of the filler and polymer matrix is a crucial factor in determining the interfacial thermal resistance. In a previous research study that made use of atomic-scale experiments, it has also been found that the degree of similarity in vibrational properties controls the heat transport across an interface.[47] To probe the underlying mechanisms for the effect of functionalisation on graphene-paraffin interfacial thermal resistance, the VDOS for pristine graphene, functionalised graphene and paraffin were calculated. The VDOS in the frequency domain was obtained by taking the Fourier transform of the velocity autocorrelation functions of atoms.[48] Specifically, the VDOS at vibrational wavenumber, $\omega$, is given by

$$P(\omega) = \frac{1}{\sqrt{2\pi}} \int e^{i\omega t} \langle \sum_{j=1}^{N} v_j(t)v_j(0) \rangle dt \qquad (3)$$

where $v_j(0)$ and $v_j(t)$ are the velocities of atom $j$ at initial time and at time $t$, respectively; and $N$ is the number of atoms. Figure 5(a) shows the VDOS of pristine graphene and paraffin. It can be seen that the overall VDOS of pristine graphene peaks at about 53 THz and 10 to 18 THz. The peak at about 53 THz represents the in-plane phonon modes and it is caused by the vibrations of covalent carbon-carbon (C-C) bonds. The peaks at lower frequencies (10 to 18 THz) indicate the out-of-plane phonon modes which arise from the flexibility of graphene in its basal plane. The VDOS of graphene agrees well with the available results in literature.[35,48,49] As shown in Figure 5(a), the VDOS of paraffin is isotropic and it has a peak around 88 THz due to the carbon-hydrogen (C-H) bonds and 20 to 45 THz due to other interactions. The frequency of C-H bonds is high because of the light hydrogen atoms. It is readily seen from Figure 5(a) that there is a small overlap between the VDOS spectra of the graphene and paraffin. To quantitatively determine the overlap of two VDOS spectra, a correlation factor $S$ is defined as[50]

$$S = \frac{\int P_p(\omega)P_g(\omega)d\omega}{\int P_p(\omega)d\omega \int P_g(\omega)d\omega} \qquad (4)$$

where $P_p$ denotes the spectra of paraffin and $P_g$ is the spectra of pristine or functionalised graphene. A small correlation factor indicates that the two spectra possess a poor overlap. For the VDOS of pristine graphene and paraffin, $S$ is 0.006. Both the visually poor VDOS overlap and small correlation factor validate the high interfacial thermal resistance between the pristine graphene and paraffin as reported by other researchers.[30,35,48]

Figure 5(b) shows the VDOS of paraffin and graphene functionalised with butyl group. The comparison reveals that their VDOS overlap on multiple peaks in both high and low frequency regions. These overlaps and a high $S$ value of 0.972 clearly explain the prominent reduction of graphene-paraffin interfacial thermal resistance induced by butyl functionalisation. As shown in Figures 5(c)-(h), when the VDOS of graphene functionalised with other functional groups are compared to that of paraffin, the overlap of VDOS peaks is found to be consistent with the trend of interfacial thermal resistance observed in Figure 4. For instance, the VDOS of methyl, phenyl and formyl functionalised graphene have moderate overlap with that of paraffin, which supports the moderately effective reduction of interfacial thermal resistance. The VDOS of carboxyl, hydroxyl and amines functionalised graphene barely overlap with that of paraffin. As a result, these functional groups were identified to be ineffective in reducing the graphene-paraffin interfacial thermal resistance. It is worth noting that the present work focused on the interfaces between the polymer and functionalised graphene without direct chemical bonding, which may be more representative of the real applications today. In the previous studies, it was suggested that forming direct chemical bonds between two interfacing materials may further increase the interfacial thermal conductance by more than one order of magnitude.[51,52]

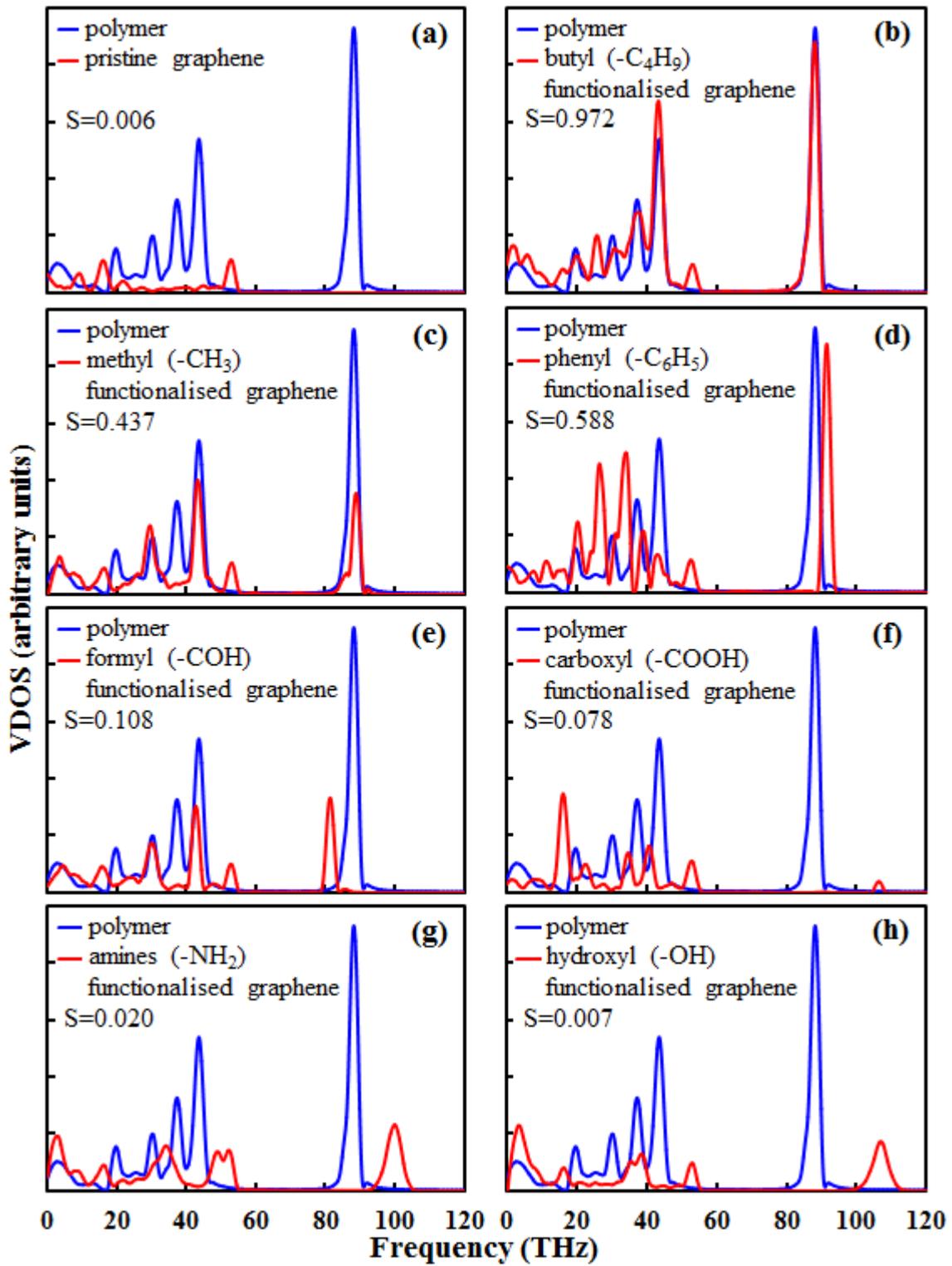

**Figure 5.** Comparison of VDOS between polymer and various types of graphene.

**3.2. Effect of Number of Graphene Layers**

Besides single-layer graphene, bilayer and multilayer graphene have also been widely used as fillers in polymeric composite TIMs for thermal management applications.[9,10] In this section, the effect of graphene layer number on the thermal transport across graphene-paraffin interfaces was investigated. For this investigation, the single-layer graphene model described above was duplicated and stacked to form 2 to 4-layer graphene with an interlayer spacing of 3.35 Å. Both pristine graphene and functionalised multilayer graphene were considered. For the functionalised models, butyl and methyl were chosen as the functional groups and the functionalisation was randomly added onto the two outer surfaces of the multilayer graphene. The coverage of functionalisation is 5.36% on each surface. The multilayer graphene models were simulated respectively in the polymeric composite system to calculate the interfacial thermal resistance. As an example, the model composed of functionalised 3-layer graphene and paraffin under RNEMD simulation is shown in Figure 6.

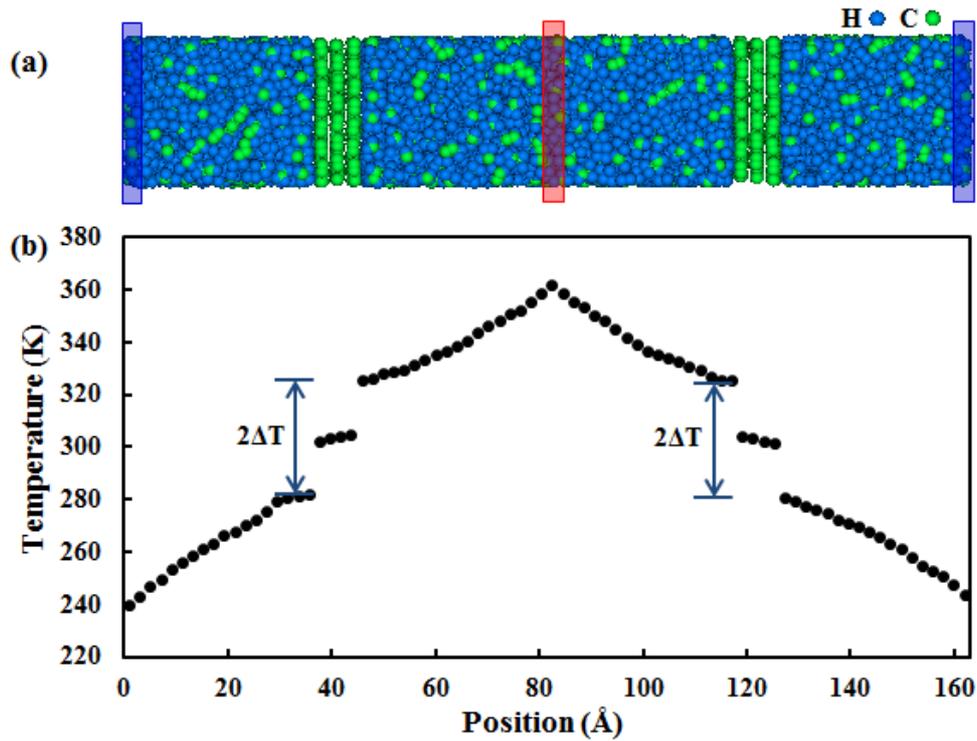

**Figure 6.** (a) Composite composed of 3-layer graphene and polymer; and (b) the established temperature gradient.

Figure 7 shows the variation of the relative interfacial thermal resistance ($R_K/R_{K0}$) with respect to the number of graphene layers. Here, $R_{K0}$ is denoted as the interfacial thermal resistance between paraffin and pristine single-layer graphene. $R_K$ is the interfacial thermal resistance between paraffin and pristine or methyl functionalised multilayer graphene. As shown in Figure 7, for both pristine and functionalised graphene, the interfacial thermal resistance roughly remains unchanged as the number of graphene layers increases, indicating the independence of the layer number.

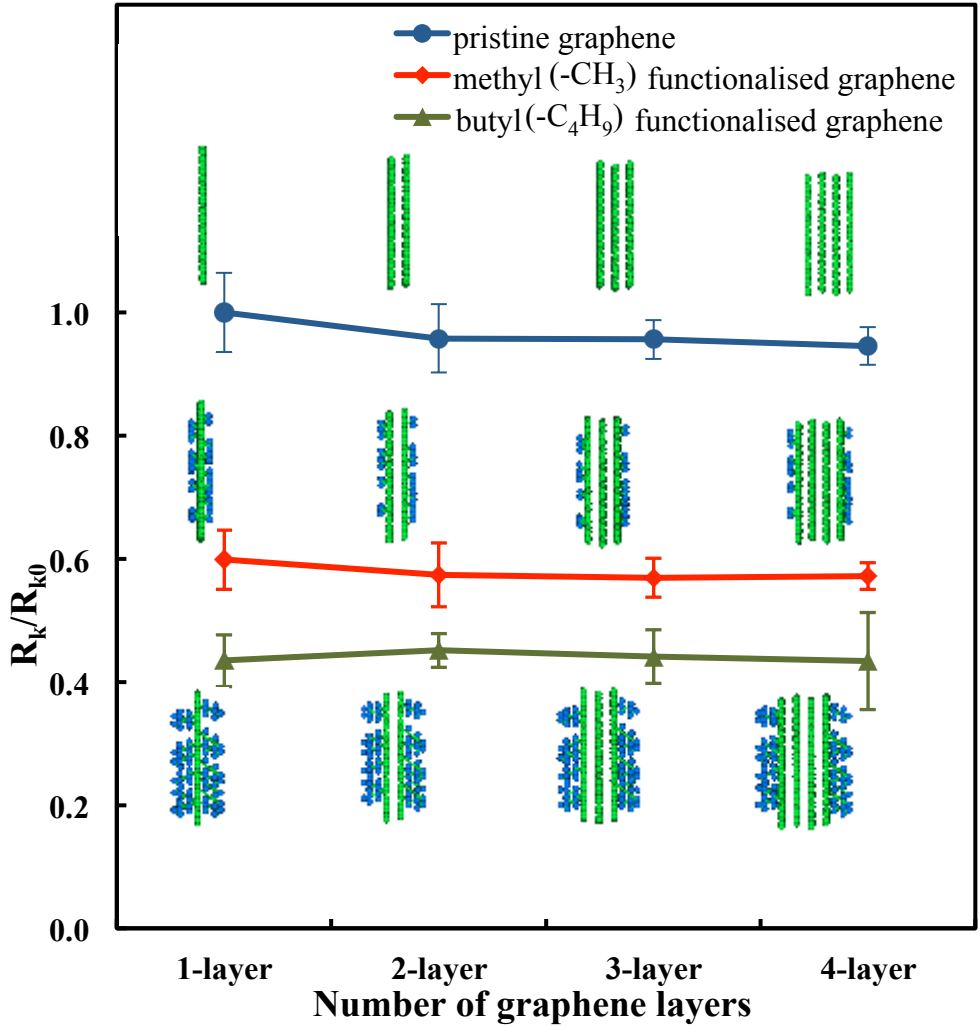

**Figure 7.** Variation of the relative interfacial thermal resistance with respect to the number of graphene layers.

Figure 8 shows the VDOS of paraffin and interfacial graphene with attached functional groups. It can be seen that the VDOS overlapping does not change significantly with different graphene layer number. This supports the layer number-insensitive interfacial thermal resistance as shown in Figure 7. These results also agree with the phenomenon observed by Hu et al.[30] Using similar MD simulation methods, they reported that the graphene-polymer interfacial thermal conductance values are similar when the number of pristine graphene layers varies between 1, 3 and 5. The results obtained here are consistent with the results published in the literature and the same conclusion is reached for functionalised graphene.

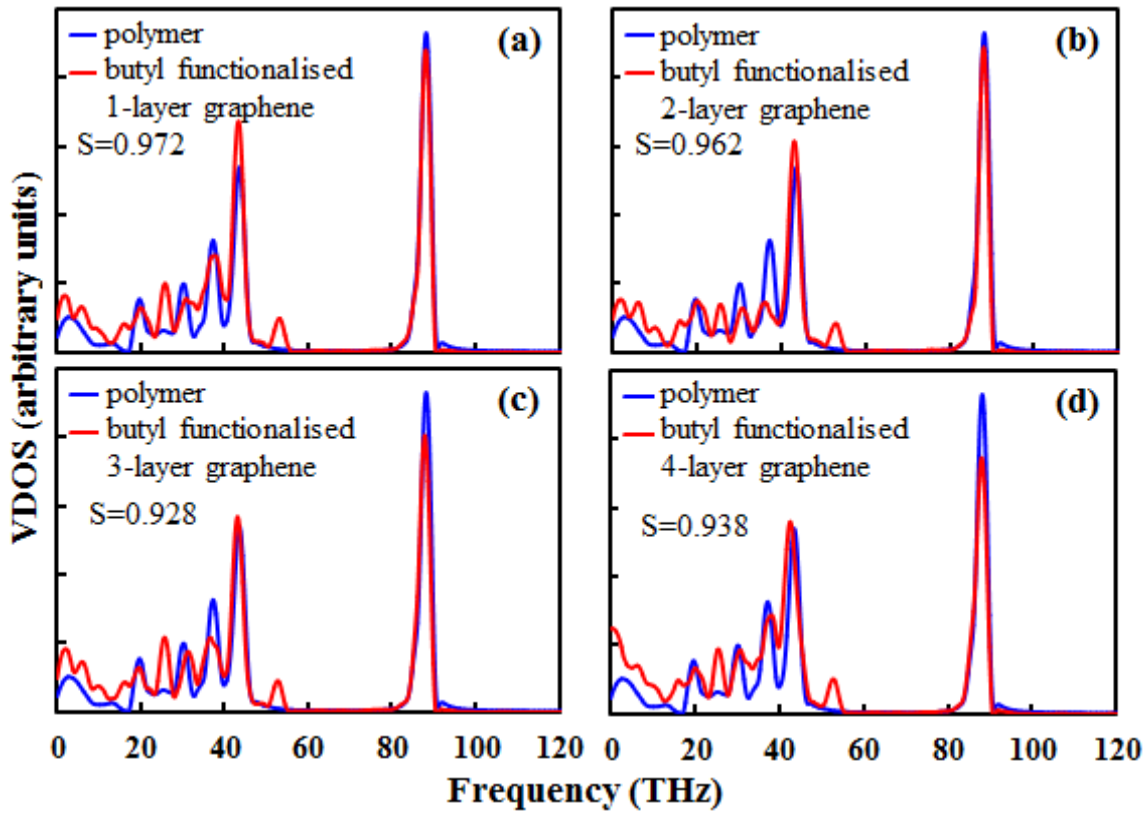

**Figure 8.** VDOS comparison between polymer and butyl functionalised interfacial graphene with respect to the number of graphene layers.

## 4. CONCLUSIONS

In summary, the thermal transport across graphene-paraffin interfaces has been investigated via MD simulations. Based on the simulation results, it is found that covalent functionalisation significantly affects the thermal transport across graphene-paraffin interfaces. The presence of functionalisation leads to a reduction of interfacial thermal resistance and this reduction is linearly dependent on the coverage. At the same coverage, butyl has been identified as the most effective functional group in reducing the interfacial thermal resistance, followed by methyl, phenyl and formyl. Other functional groups including carboxyl, hydroxyl and amines have been found to have a negligible effect on the graphene-paraffin interfacial thermal transport. The effects of different functional groups are explained by analysing the VDOS overlap of the graphene and paraffin. Different types of functionalised graphene possess different degrees of VDOS overlap with paraffin, which result in different interfacial thermal resistances. The number of graphene layers has been found to have a negligible effect on the graphene-paraffin interfacial thermal transport. Our simulation results provide useful guidelines on the selection of appropriate functionalisation for enhancing graphene-polymer interfacial thermal transport. The findings herein should be useful for the future development of high performance graphene-polymer composite TIMs.

**ACKNOWLEDGMENTS**

Y. Wang acknowledges the financial support on his PhD study from Australian Government via Australian Postgraduate Awards Scheme. The computational support provided by Intersect Australia Ltd (INTERSECT) and National Computing Infrastructure (NCI) is gratefully acknowledged.

**Table of Contents (TOC) - Graphic**

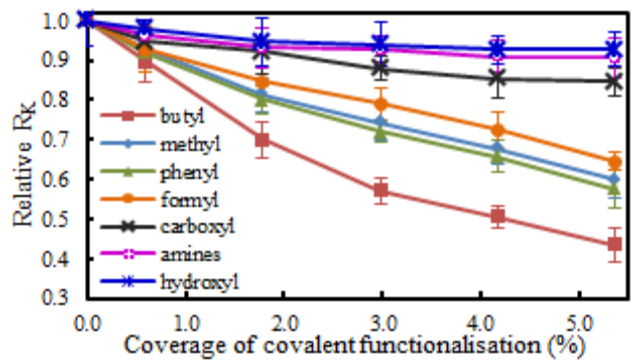